\documentclass[aps,pra,twocolumn,nofootinbib,a4paper,floatfix,10pt]{revtex4-2}

\usepackage{color}
\usepackage{bm}
\usepackage{amsmath,amssymb}
\usepackage{amsfonts}
\usepackage{dsfont}
\usepackage{graphicx}
\usepackage{float}
\usepackage{subfigure}
\usepackage{verbatim}
\usepackage{amsfonts}
\usepackage{dcolumn}
\usepackage{bm}
\usepackage{color}
\usepackage[dvipsnames]{xcolor}
\usepackage{listings}
\definecolor{zaffre}{rgb}{0.0, 0.08, 0.66}
\usepackage[colorlinks=true,urlcolor=zaffre,citecolor=zaffre,linkcolor=zaffre]{hyperref}
\usepackage{mathrsfs}
\usepackage{filecontents}
\usepackage{mathtools}
\usepackage{times} 

\begin{document}

\title{Simple and loss-tolerant free-space QKD using a squeezed laser}

\author{Nedasadat Hosseinidehaj}  \email{nedahsn@gmail.com}
\affiliation{Centre for Quantum Computation and Communication Technology, School of Mathematics and Physics, University of Queensland, St Lucia, Queensland 4072, Australia}

\author{Matthew S. Winnel}
\affiliation{Centre for Quantum Computation and Communication Technology, School of Mathematics and Physics, University of Queensland, St Lucia, Queensland 4072, Australia}

\author{Timothy C. Ralph}
\affiliation{Centre for Quantum Computation and Communication Technology, School of Mathematics and Physics, University of Queensland, St Lucia, Queensland 4072, Australia}
\date{\today}

\begin{abstract}

We consider a continuous-variable (CV) quantum key distribution (QKD) protocol over free-space channels, which is simpler and more robust than typical CV QKD protocols. It uses a bright laser, squeezed and modulated in the amplitude quadrature, and self-homodyne detection.  
We consider a scenario, where the line of sight is classically monitored to detect active eavesdroppers, so that we can assume a passive eavesdropper.  Under this assumption, we analyse security of the QKD protocol in the composable finite-size regime. Proper modulation of the squeezed laser to the shot-noise level can completely eliminate information leakage to the eavesdropper and also eliminate the turbulence-induced noise of the channel in the amplitude quadrature. Under these conditions, estimation of the eavesdropper's information is no longer required. 
The protocol is extremely robust to modulation noise and highly loss-tolerant, which makes it a promising candidate for satellite-based continuous-variable quantum communication.     

\end{abstract}

\maketitle

\section{Introduction}\label{sec1}

\begin{figure}
      {\includegraphics[width=2.4 in]{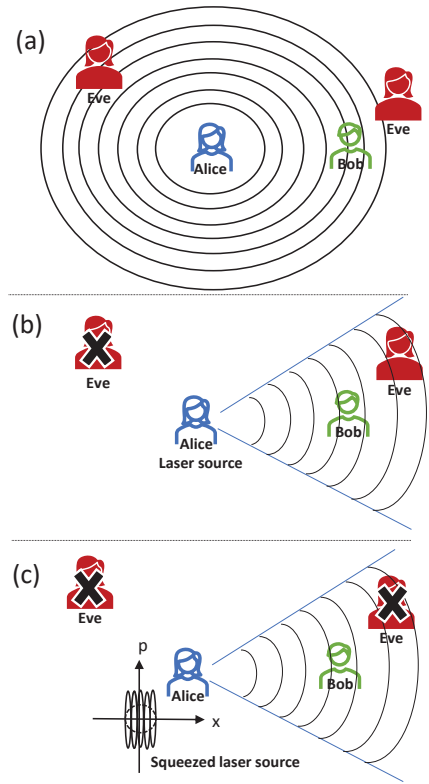}}
   \caption{A schematic representation of communication scenarios between Alice and Bob in the presence of an eavesdropper, Eve: (a) current wireless communication, which is only secure given computational limitations of Eve; (b) communication using lasercomm; and (c) the QKD protocol we consider in this paper using a squeezed laser, which can completely eliminate information leakage to Eve.}\label{QKD}
\end{figure}

Current wireless communication systems are omni-directional and so are easy to eavesdrop upon (see Fig.~\ref{QKD}(a)). Public-key cryptography can be used to secure such transmissions, offering security via assumptions about the computational power of malicious
eavesdroppers. These assumptions are called into question by possible future advances in computational power, in particular the advent of large-scale quantum computation \cite{QCQI}. Given this, the security of current communications cannot be guaranteed indefinitely as they might be stored and decrypted in the future when the required level of quantum processing becomes available. Laser communications (lasercomm) can improve security in certain circumstances via its greater directionality. Nevertheless eavesdropping is still possible due to beam diffraction (see Fig.~\ref{QKD}(b)). Here we propose a simple extension to lasercomm using techniques from quantum key distribution \cite{Scarani.et.al.RVP.09, Pirandola.et.al.arxiv.19} and recent observations about information leakage \cite{leakage-elimination}, which eliminates these eavesdroppers  (see Fig.~\ref{QKD}(c)).

Quantum key distribution (QKD) allows two
trusted parties, Alice and Bob, to share a secure key, unknown to a potential eavesdropper, Eve.  In contrast to current classical cryptography, QKD provides information-theoretic security \cite{Scarani.et.al.RVP.09,Xu.et.al.arxiv.19,Pirandola.et.al.arxiv.19}. Although QKD started with discrete-variable quantum systems \cite{Bennett.Brassard.IEEE.84,Ekert.PRL.91}, it has been extended to continuous-variable (CV) systems \cite{Ralph.PRA.99,Hillery.PRA.00,Reid.PRA.00,RR2003}. In the former, information is encoded in discrete degrees of
freedom of a single photon, with the detection realised by single-photon detectors. While, in the latter, information is
encoded in continuous degrees of freedom of light, and detection is realised
by off-the-shelf homodyne detectors, offering greater compatibility with current optical telecommunication networks. 

In a typical lasercomm protocol information is encoded via amplitude modulation of the laser beam and read out via direct detection, also known as self-homodyne. Similarly, in this work we propose a CV QKD protocol based on amplitude modulation of a {\it squeezed} laser with read out also via direct detection. In contrast, in a typical CV QKD protocol information is encoded in both amplitude and phase quadratures of the light while the detection is performed by either homodyne or heterodyne detectors, requiring the use of a separate local oscillator \cite{Garcia-Patron.PhD.07,Weedbrook.et.al.RVP.12,Diamanti.Leverrier.Entropy.15,Pirandola.et.al.arxiv.19}. Our simplification comes by specifically considering  free-space channels and hence limiting the eavesdropper to only gathering the lost light - i.e. a passive attack. Whilst this is not the most general attack, we argue it is a reasonable restriction given plausible technical capabilities of Eve. Given this restriction we make a full, composable finite-key security analysis of our system and find it is robust to loss, turbulence and excess noise of the source laser.

Free-space channels are flexible in terms of infrastructure establishment and feasibility of communication with moving objects. They also provide the possibility of long-distance quantum communication via orbiting satellites. The key disadvantage of free-space quantum communications is, however, atmospheric attenuation and turbulence noise \cite{Hosseini.et.al.IEEE.19, satellite-2017-1, satellite-2017-2}. Atmospheric turbulence causes a random variation of channel transmissivity in time \cite{Semenov.Vogel.PRA.09, Vasylyev.et.al.PRL.12, Vasylyev.et.al.PRL.16, Vasylyev.et.al.PRA.17, Vasylyev.et.al.PRA.18}. This transmissivity fluctuation introduces extra noise on CV QKD systems, which reduces the secret key rate, and even renders the communication insecure in the presence of strong turbulence \cite{Hosseini.Malaney.PRA1.15, Dong-PRA-10, Usenko-NJP-12, Hosseini.Malaney.ICC.15, Hosseini.Malaney.PRA2.15, Hosseini.Malaney.GLOBECOM.16, Hosseini.Malaney.QIC.17, fast-fading, NJP-2018, cluster-2019, Neda-Nathan-Tim-freespace}. It is thus of considerable significance that reasonable restrictions on Eve can lead to a far simpler and more robust system.

\section{The model}

We analyse a CV QKD protocol using a squeezed laser over a free-space channel using modulation in only the amplitude quadrature and direct, self-homodyne detection (see Fig.~\ref{QKD}(c)). 
A proper modulation of the squeezed states leads to zero turbulence-induced noise of the free-space channel in the amplitude quadrature as well as zero information leakage to the eavesdropper. The protocol does not require estimation of the eavesdropper's information, as information leakage is zero. It is also highly robust against modulation imperfections (i.e. modulation noise), and can tolerate high values of channel loss.  

Firstly, in our QKD protocol we assume the trusted parties, Alice and Bob, are able to classically monitor the line of sight, so that any active presence of an eavesdropper (Eve) in the line of sight can be detected, and if there is any, the protocol will be aborted. Whilst an omnipotent eavesdropper might deceive Alice and Bob, the technologies required seem well beyond current capabilities. As a result, any active eavesdropping attacks in the line of sight will be prevented, and Eve will be limited to only passive attacks.

Secondly, in our QKD protocol we will exploit squeezed quantum states for information carriers, similar to the CV protocol of \cite{leakage-elimination}, where by properly encoding information into a Gaussian modulation of squeezed states (squeezed in a single quadrature and modulated in the squeezing direction), one can completely and deterministically eliminate information leakage to Eve in a reverse reconciliation (RR) scenario \cite{RR2003}, if the channel is pure loss.  The necessary condition for leakage elimination is that the variance of the average input state in the modulation direction has to be the shot-noise variance. Unlike \cite{leakage-elimination}, we consider bright squeezed states such that the modulation can be read out via direct detection of the light.

Since a pure-loss channel can be considered as a passive eavesdropping attack, by the Gaussian modulation of the squeezed laser to the shot-noise level and limiting Eve to only passive attacks, we will completely and deterministically eliminate information leakage to Eve in a RR scenario (i.e., obtain zero Holevo information) over free-space channels. As a result, estimation (or upper bounding) of Eve's information is no longer required in this protocol. Such a shot-noise modulation can also eliminate the channel-fluctuation noise in the information-carrying quadrature.



Because the protocol uses squeezed bright beams, where the information is only encoded in the amplitude quadrature, self-homodyne detection at Bob's station with no need for a separate local oscillator will suffice to measure the amplitude quadrature. This will significantly simplify the experimental realisation of the protocol. 

We further investigate the effect of modulation imperfections. We consider the case where the variance of the average input state in the squeezing direction is not exactly the shot noise. In fact, we consider some amount of trusted preparation noise on top of the shot noise in Alice's side. We show that this type of practical imperfection leads to some information leakage to Eve, where we do the security analysis in the composable finite-size regime. However, the amount of leakage is sufficiently small, so that its effect on the secret key rate is negligible. 


\section{Gaussian-modulation squeezed-state protocol}

In a prepare-and-measure scheme, Alice generates a random real variable $a_q$, drawn from a Gaussian distribution of variance $V_{\rm sig}$ and zero mean. Alice prepares bright squeezed states with the squeezing in the amplitude $\hat q$ quadrature, where the variance of the squeezed quadrature is $V_{\rm sqz}$. The squeezed states are then modulated (displaced) by an amount $a_q$ in the direction of the $\hat q$ quadrature. The variance of the average Gaussian state after the modulator is $V_{\rm sqz}+V_{\rm sig}$ in the amplitude $\hat q$ quadrature and $1/V_{\rm sqz}$ in the phase $\hat p$ quadrature. We consider the case where the variance of the squeezed quadrature after the modulation is equal to the shot-noise variance, i.e.,  $V_{\rm sqz}+V_{\rm sig}=1$ \cite{leakage-elimination}. The squeezed states are then sent through a free-space channel to Bob, who directly measures the amplitude $\hat q$ quadrature with self-homodyne detection. 

In contrast to a fiber link with a fixed transmissivity, the transmissivity, $\eta$, of a free-space channel fluctuates in time due to atmospheric turbulence. Such fading channels can be characterized by a probability distribution $p(\eta)$ \cite{Semenov.Vogel.PRA.09,Vasylyev.et.al.PRL.12,Vasylyev.et.al.PRL.16,Vasylyev.et.al.PRA.17,Vasylyev.et.al.PRA.18}. A fading channel can be decomposed into a set of sub-channels, for which the transmissivity is relatively stable. Each sub-channel ${\eta_i}$ occurs with probability $p_i$ so that $\sum\nolimits_{i} {p_i} = 1 $ or $\int_{0}^{\eta_{\rm max}} { p(\eta )} d\eta = 1$ for a continuous probability distribution, where $\eta_{\rm max}$ is the maximum realisable transmissivity.  

In order to analyse the security of the CV QKD protocol, we consider the equivalent entanglement-based scheme. Alice first prepares a symmetric two-mode squeezed vacuum state of quadrature variance $V$. One mode is kept by Alice (to be later measured via homodyne detection in the $\hat q$ quadrature), while the second mode is squeezed (in Alice's side) in the $\hat q$ quadrature with the squeezing parameter $r_e$. 
The initial entangled state, prepared on Alice's side, is given by the following covariance matrix,
\begin{equation}\label{sqz-initial-state-Gaussian}
\begin{array}{l}
{{\bf{M}}_{\rm AB_0}} {=} 
\left[ {\begin{array}{*{20}{c}}
a_q&0&{c_q }&0\\
0&a_p&0&{c_p }\\
{c_q }&0&{b_q}&0\\
0&{c_p }&0&{b_p}
\end{array}} \right], \left\{ {\begin{array}{*{20}{c}}
{{a_q} = {a_p} = V},\\
{{b_{q}} = V e^{ - 2r_e}, \,\,\, {b_p} = {V e^{2r_e}}},\\
{{c_q} = {e^{ - r_e}}\sqrt {{V^2} {-} 1}},\\
{{c_p} = {- {e^{  r_e}}\sqrt {{V^2} {-} 1}}}.
\end{array}} \right.
\end{array}
\end{equation}
Note that in order for the prepare-and-measure scheme to be equivalent with the entanglement-based scheme we need to have $V_{\rm sqz}+V_{\rm sig} = {V{e^{ - 2r_e}}}$ and $1/V_{\rm sqz} = {V{e^{  2r_e}}}$. 

\subsection{Eavesdropping assumptions}
In practice, it would be very challenging for Eve to make an active attack (for instance an entangling cloner attack \cite{Entangling-cloner-2008}) over a free-space channel. This is because Alice and Bob can classically monitor the line of sight for any active presence of Eve, and abort the protocol if they find anything. Alternatively, Eve could perform a shine-on attack, by using an entangled state and passively adding extra noise onto Bob's detector. Again, this attack will be very challenging for Eve in a self-homodyne detection scenario with the phase of the local oscillator being  random, and also the line-of-sight being monitored. However, even if Eve can invisibly add extra noise, it will be identified by Alice and Bob in unexpected deviations from shot-noise at Bob's station and when they estimate a signal-to-noise ratio (SNR) lower than that they expect from the light-collection attack. 
Thus, with no reduction in practical security, Eve's attack over free-space channels can be restricted to a passive attack, which is the same as a beam-splitter attack. In a passive collective attack, Eve collects the light lost in the transmission, and stores the quantum states in her quantum memory to be collectively measured later. 

\section{Finite-size security analysis}

The Wigner function of Alice and Bob's ensemble-average state (over all sub-channels) at the output of a free-space channel with fluctuating transmissivity $\eta$ is the sum of the Wigner functions of the states after individual sub-channels weighted by sub-channel probabilities \cite{Neda-Nathan-Tim-freespace}. Alice and Bob's state is Gaussian after the realisation of each sub-channel, however, Alice and Bob's ensemble-average state is a non-Gaussian mixture of Gaussian states obtained from individual sub-channels.

In an RR scenario, Eve's information, upper bounded by the Holevo information in a collective attack, is given by $\chi(b{:}E)=\mathcal{S}(\rho_{E})-\mathcal{S}(\rho_{E|b})$, where $\mathcal{S}(\rho)$ is the von Neumann entropy of the quantum state $\rho$. The security is analysed based on the purification assumption, i.e., the assumption that Alice and Bob's quantum state $\rho_{AB}$ is purified by Eve's quantum state $\rho_E$.
This results in $\mathcal{S}(\rho_E)=\mathcal{S}(\rho_{AB})$, and $\mathcal{S}(\rho_{E|b})=\mathcal{S}(\rho_{A|b})$. Thus, Eve's Holevo information is given by $\chi(b{:}E)=\mathcal{S}(\rho_{AB})-\mathcal{S}(\rho_{A|b})$. Note that $\rho_{AB}$ is non-Gaussian for a free-space channel, however, according to the optimality of Gaussian attacks \cite{Wolf.Giedke.Cirac.PRL.06, Navascues.Grosshans.PRL.06, Garcia-Patron.Cerf.PRL.06}, $\chi(b{:}E)$ can be maximised if calculated based on the covariance matrix of $\rho_{AB}$. 
The covariance matrix of Alice and Bob's ensemble-average state is given by
\begin{equation}\label{effective-AliceBob}
\begin{array}{l}

{{\bf{M}}_{A{B}}} {=} 
\left[ {\begin{array}{*{20}{c}}
a_q&0&{c'_q }&0\\
0&a_p&0&{c'_p }\\
{c'_q }&0&{b'_q}&0\\
0&{c'_p }&0&{b'_p}
\end{array}} \right], {\rm{where}}
\\
\\
b'_q = \eta_f b_q + 1 - \eta_f + {\rm{Var}}(\sqrt \eta  )(b_q - 1) 
\\
b'_p = \eta_f b_p + 1 - \eta_f + {\rm{Var}}(\sqrt \eta  )(b_p - 1) 
\\
c'_q = \sqrt{\eta_f} \, c_q , \,\,\, c'_p = \sqrt{\eta_f} \, c_p, {\rm{where}}\\
\\
\eta_f = {\left\langle {\sqrt{\eta}} \right\rangle}^2, \,\,\,
\rm{Var}(\sqrt \eta  ) = \left\langle \eta  \right\rangle  - {\left\langle {\sqrt \eta  } \right\rangle ^2},
\end{array}
\end{equation}
where $ \langle \eta  \rangle = \int_{0}^{\eta_{\rm max}} { \eta p(\eta )} d\eta $, and $ \langle \sqrt{\eta}  \rangle = \int_{0}^{\eta_{\rm max}} { \sqrt{\eta} p(\eta )} d\eta $. Thus, Eve's effective passive attack can be considered as a beam-splitter attack with the beam-splitter transmissivity $\eta_f$. 

According to Eq.~(\ref{effective-AliceBob}), a fading channel is equivalent with a fixed-transmissivity channel with transmissivity $\eta_f $ and an extra non-Gaussian noise of ${\rm{Var}}(\sqrt{\eta}) (b_{q(p)}-1)$ \cite{Usenko-NJP-12, cluster-2019, Neda-Nathan-Tim-freespace}. This noise depends on the channel fluctuation variance ${\rm{Var}}(\sqrt{\eta})$ and the input variance to the channel $b_{q(p)}$. When Eve's attack is considered passive, it means that the channel fluctuation is not under Eve's control. This means that the fluctuation noise ${\rm{Var}}(\sqrt{\eta}) (b_{q(p)}-1)$ is not accessible to Eve for the purification, and hence the fluctuation noise should be considered as a trusted noise in Bob's side. On the other hand, having a trusted noise in Bob's side in a RR scenario decreases Eve's information \cite{trustednoise}. Hence, in calculating Eve's information from the passive attack we consider the (trusted) fluctuation noise to be zero as this can only overestimate Eve's information. 

Note that we can also use Eve and Bob's covariance matrix to calculate Eve's Holevo information from the passive attack, and obtain the same result as that from the purification assumption as discussed above (see Appendix~\ref{AppendixA} for more details on the purification assumption). 
As a result of Eve's passive attack, the covariance matrix of Eve's ensemble-average state is given by
\begin{equation}\label{E-CM-channel-sqz}
\begin{array}{l}
{{\bf{M}}_E} = 
\left[ {\begin{array}{*{20}{c}}
{(1 {-} \eta_f )b_q   {+} \eta_f }&0\\
0&{(1 {-} \eta_f )b_p   {+} \eta_f }
\end{array}} \right],
\end{array}
\end{equation}
The covariance matrix of Eve's system conditioned on Bob's homodyne detection (with efficiency $\eta_B$ and electronic noise $\nu_B$) is given by ${{\bf{M}}_{E\left| {B'} \right.}} {=} {{\bf{M}}_E} {-} {{\bf{M}}_{E{B'}}}{\left( {{\bf{X}}} {{\bf{M}}_{B'}} {{\bf{X}}}  \right)^{ \rm MP}} {\bf{M}} _{E{B'}}^T$, where ${{\bf{X}}} = {\rm diag}(1,0)$, MP stands for the
Moore-Penrose pseudoinverse of a matrix, and we have  
\begin{equation}\label{EB-CM-channel-sqz}
\begin{array}{l}
{{\bf{M}}_{B'}}  = {\rm diag}(V_{Bq},V_{Bp}),\,\, {\rm where} \,\, \\
\\
V_{Bq} = {{\eta _B}\left[ { \eta_f   b_q  {+} 1 {-}  \eta_f   } \right] {+} (1 {-} {\eta _B})\upsilon },\\
V_{Bp} = {{\eta _B}\left[ { \eta_f   b_p  {+} 1 {-}  \eta_f   } \right] {+} (1 {-} {\eta _B})\upsilon },\,\, {\rm where} \,\, \\
\\
\upsilon = 1+\frac{ \nu_B}{1-\eta_B},\,\, {\rm and} \\
\\
{{\bf{M}}_{E{B'}}} = {\rm diag}(C_{EBq},C_{EBp}) ,\,\, {\rm where} \,\, \\
\\
C_{EBq} = {\sqrt {{\eta _B}}  {\sqrt {\eta_f (1 - \eta_f )} }  \left[ {1 - b_q  } \right]},\\
C_{EBp} = {\sqrt {{\eta _B}}  {\sqrt {\eta_f (1 - \eta_f )} }  \left[ {1 - b_p } \right]}.
\end{array}
\end{equation}
Note that in reality for Bob's quadrature variance (after the detection) we have $V_{Bq} = {{\eta _B} b'_q  {+} (1 {-} {\eta _B})\upsilon } $ and $V_{Bp} = {{\eta _B} b'_p  {+} (1 {-} {\eta _B})\upsilon } $. But, since as noted earlier, the (trusted) fluctuation noise ${\rm{Var}}(\sqrt{\eta}) (b_{q(p)}-1)$ in Bob's side decreases Eve's information, we assume the fluctuation noise is zero in $V_{Bq}$ and $V_{Bp}$ of Eq.~(\ref{EB-CM-channel-sqz}) for the security analysis. 

According to the protocol, Alice has to make sure that the beam leaving her lab in the prepare-and-measure scheme has exactly the shot-noise variance in the $\hat q$ quadrature. It means that the beam leaving Alice's lab in the entanglement-based scheme also needs to have the shot-noise variance in the $\hat q$ quadrature, i.e., $b_q = 1 $. As a result, according to ${{\bf{M}}_{EB}}$ in Eq.~(\ref{EB-CM-channel-sqz}), there is no correlation between Eve and Bob in the $\hat q$ quadrature ($C_{EBq} = 0 $), i.e, the quadrature that contains the key information. Hence, there is no information leakage to Eve during the quantum communication part in a RR scenario, i.e., we have the Holevo information $\chi({b{:}E})= \mathcal{S}({\bf{M}}_{E}){-}\mathcal{S}({{\bf{M}}_{E\left| {B'} \right.}}) = 0$. 

The shot-noise modulation in the $\hat q$ quadrature has another advantage of eliminating the fluctuation-induced noise of a free-space channel. Bob's variance in the $\hat q$ quadrature (before the detection) is given by $b'_q {=} \eta_f b_q + 1 - \eta_f + {\rm{Var}}(\sqrt \eta  )(b_q - 1) $.
When we have $b_q = 1 $, the fluctuation-induced noise of the channel in the $\hat q$ quadrature, i.e. ${\rm{Var}}(\sqrt \eta  )(b_q - 1)$, will become zero, and Bob's variance will also be the shot noise, i.e., $b'_q {=} 1$. 

Since having $b_q = 1 $ results in no information leakage to Eve during the quantum communication, i.e. $\chi({b{:}E}){=}0$, we do not need to estimate (or upper bound) Eve's information. However, the transmissivity of the channel needs to be estimated in order to estimate the SNR of the channel, which will be used to choose the most efficient error-correcting code rate for the error-correction step. This means we are  still required to reveal a subset of data for SNR estimation.

Note that in this protocol, $\chi(b{:}E)=0$ does not mean that Eve and Bob's quantum systems are not correlated because Eve and Bob still remain correlated in the phase $\hat p$ quadrature ($C_{EBq} \neq 0$). However, this correlation is irrelevant to the security of the protocol because the key information is only encoded in the $\hat q$ quadrature. In fact, $\chi(b{:}E)=0$ means that Bob's measurement outcomes are uncorrelated with Eve's quantum system $E$ before the error correction. However, in the error-correction procedure, classical information $C$ of size $l_{\rm EC}$ (i.e., the size of the syndrome of Bob's string sent to Alice in a RR scenario) will be revealed by the trusted parties. 
In the privacy amplification step, Alice and Bob have to discard the leakage during the error correction.

Based on the leftover hash lemma \cite{lemma1,lemma2}, the number of approximately secure bits, $\ell$, that can be extracted from the raw key should be slightly smaller than the smooth min-entropy of Bob's string $b$ conditioned on Eve's system $E'$ (which characterizes Eve's quantum state $E$, as well as the public classical variable $C$ leaked during the QKD protocol), denoted by $H_{\min}^{\epsilon_{\rm sm}}(b^{N'}|E')$ \cite{lemma1}, i.e., we have $\ell \le  H_{\min}^{\epsilon_{\rm sm}}(b^{N'}|E') {-} 2\log_2 (\frac{1}{{2\bar \epsilon }}) $, where $\bar \epsilon$ comes from the leftover hash lemma. Note that $N'$ indicates the length of Bob's string $b$ after the SNR estimation.
The chain rule for the smooth min-entropy \cite{Finite-size-Leverrier-2015} gives $H_{\min}^{\epsilon_{\rm sm}}(b^{N'}|E')=H_{\min}^{\epsilon_{\rm sm}}(b^{N'}|EC) \ge  H_{\min}^{\epsilon_{\rm sm}}(b^{N'}|E) - \log_2|C|$, where $\log_2|C|=l_{\rm EC}$, with $l_{\rm EC}$ the size of data leakage during the error correction, which can be given by $l_{\rm EC} = N' [H(b) - \beta I(a{:}b)]$ \cite{Finite-size-Leverrier-2015,Finite-size-Furrer,Finite-size-Lupo-MDI}, where $H(b)$ is Bob's Shannon entropy and $\beta$ is the reconciliation efficiency.
In order to calculate the length $\ell$ of the final key which is $\epsilon$-secure ($\epsilon {=} 2\epsilon_{\rm sm}{+}\bar \epsilon {+}\epsilon_{\rm PE}{+}\epsilon_{\rm cor}$ \cite{Finite-size-Leverrier-2015,Finite-size-Lupo-MDI}), the conditional smooth min-entropy $H_{\min}^{\epsilon_{\rm sm}}(b^{N'}|E)$ has to be lower bounded when the protocol did not abort.
Under the assumption of independent and identically distributed (i.i.d) attacks such as collective attacks (which we consider here), the asymptotic equipartition property \cite{Finite-size-Leverrier-2015, Marco-thesis, Marco} can be utilized to lower bound the conditional smooth min-entropy with the conditional von Neumann entropy. Explicitly, we have $ H_{\min }^{\epsilon_{\rm sm}} (b^{N'}\left| E \right.) \ge N' \mathcal{S}(b\left| E \right.) - \sqrt{ N'}  \Delta_{\rm AEP}$ \cite{Finite-size-Leverrier-2015,Finite-size-Lupo-MDI}, where $\Delta_{\rm AEP} = (d{+}1)^2{+}4(d{+}1)\sqrt{\log_2({2{/}\epsilon_{\rm sm}^2})} + 2\log_2({2}{/}({\epsilon^2 \epsilon_{\rm sm}}))  {+} 4{\epsilon_{\rm sm}d}{/}{(\epsilon \sqrt{ N'}) }$ with $d$ the discretization parameter, and $\mathcal{S}(b\left| E \right)$ the conditional von Neumann entropy, which is given by $\mathcal{S}(b|E)=H(b)- \chi^{\epsilon_{\rm PE}}(b{:}E)$. Eve's information on Bob's string $b$ is upper bounded by Holevo information $\chi^{\epsilon_{\rm PE}}(b{:}E)$, except with probability $\epsilon_{\rm PE}$. Recall again that having $b_q = 1 $, we do not need to estimate $\chi^{\epsilon_{\rm PE}}(b{:}E)$, as it is exactly zero. Therefore, the secret key length is given by $\ell {=} N' \beta I(a{:}b) {-} \sqrt{ N'}  \Delta_{\rm AEP}  {-} 2\log_2 (\frac{1}{{2\bar \epsilon }}) $, and the secret key rate is given by $K = \ell/N $. Note that the mutual information is given by $I(a{:}b) = \frac{1}{2} {\rm log_2} \frac{a_q}{a_q-{[\eta_B \, c'^2_q]}/[{{\eta _B} b'_q  {+} (1 {-} {\eta _B})\upsilon }]}$.
\subsection{Modulation noise}

Now we investigate the discrepancies between the ideal protocol and its practical implementations in terms of the modulation.
We consider the case where the prepared state on Alice's side does not have the exact shot-noise variance in the modulation direction. More precisely, we assume some preparation noise $\xi$ in Alice's side, which is assumed to be trusted in the case of a passive eavesdropper. 
For the aim of numerical simulation, we assume $b_q=1$ and $\xi = 0.02$. 
Eve's information can still be calculated using Eve and Bob's covariance matrices in Eqs.~(\ref{E-CM-channel-sqz}) and (\ref{EB-CM-channel-sqz}), where the term $b_{q(p)}$ in Eqs.~(\ref{E-CM-channel-sqz}) and (\ref{EB-CM-channel-sqz}) should now be replaced by $b_{q(p)}+\xi$. In this non-ideal modulation case, we have $C_{EBq} \neq 0$ in Eq.~(\ref{EB-CM-channel-sqz}) due to $b_q + \xi > 1 $, which means the preparation noise on top of the shot noise leads to information leakage (i.e. $\chi^{\epsilon_{\rm PE}}(b{:}E) \neq  0$), and the secret key length is given by $\ell {=} N' \beta I(a{:}b) {-} N' \chi^{\epsilon_{\rm PE}}(b{:}E) {-} \sqrt{ N'}  \Delta_{\rm AEP}  {-} 2\log_2 (\frac{1}{{2\bar \epsilon }})$. Note that in the presence of modulation noise, Eve's information can also be calculated from the purification assumption, i.e., using Alice and Bob's covariance matrix while assuming the (trusted) fluctuation noise is zero (see Appendix~\ref{AppendixA}).

In Fig.~\ref{keyrate}, the finite-size key rate of the squeezed-state protocol is shown as a function of channel loss under the assumption of passive collective attacks. 
For the sake of comparison, we also show the finite-size key rate of a CV QKD protocol using coherent states, under the assumption of passive attacks. Similar to the squeezed-state protocol, it uses Gaussian modulation in only the amplitude quadrature, and direct detection of the amplitude quadrature \cite{UniDcoherent}. For this protocol, the Holevo information can still be calculated using the same method as discussed for the squeezed-state protocol, where now we should set $r_e=-{\rm{ln}}( \sqrt{V})$ and $V_{\rm{sqz}}=1$. As can be seen
from Fig.~\ref{keyrate}, for losses above 4 dB, the squeezed-state protocol
outperforms the coherent-state protocol under the assumption
of passive collective attacks. The squeezed-state protocol can achieve reasonable key rates for losses more than 4 times that of the coherent-state protocol.


As can be seen from Fig.~\ref{Holevo} (showing Holevo information $\chi^{\epsilon_{\rm PE}}(b{:}E)$ as a function of channel loss for the squeezed state protocol of Fig.~\ref{keyrate}), the amount of leakage due to the preparation noise is sufficiently small such that it only has a negligible effect on the secret key rate shown in Fig.~\ref{keyrate}. For instance, even for a large preparation noise of $\xi = 0.02$, we have $\chi^{\epsilon_{\rm PE}}({b{:}E}) < 5 \times 10^{-5}$ for the given values of squeezing (very small  compared to the secret key rate shown in Fig.~\ref{keyrate}).  
Also, Fig.~\ref{Holevo} shows that Eve's information is maximised for the channel loss of around 3 dB and then reduced with increasing loss. Note that in the case of modulation noise, where there is some information leakage to Eve, parameter estimation of channel transmissivity and preparation noise is required to upper bound Eve's information $\chi^{\epsilon_{\rm PE}}({b{:}E})$ (see Appendix~\ref{AppendixB} for more details). Note that the squeezed-state protocol is very robust to error bars of estimators, such that the difference between $\chi^{\epsilon_{\rm PE}}({b{:}E})$ (i.e., Holevo information considering the error bars due to the finite-size effects) and Holevo information given the perfect estimation of channel parameters (i.e, the asymptotic case) is negligible. Note that this is not the case for the  coherent-state protocol shown in Fig.~\ref{keyrate}.

\begin{figure}[t]
    \begin{center}
      {\includegraphics[width=3.4 in, height=2.7 in]{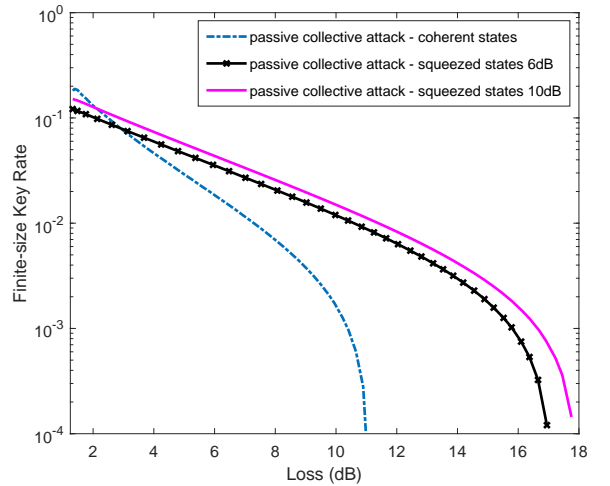}}
   \caption{Finite-size key rate as a function of channel loss (dB), where ${\rm Loss (dB)} = -10 \log_{10}{\eta_f }$, for the coherent-state protocol 
   (blue dot-dashed line) and 
   the squeezed-state protocol secure against passive collective attacks, with 6 dB (black line) and 10 dB (magenta line) squeezing, where ${\rm Squeezing (dB)} = -10 \log_{10}{V_{\rm sqz} }$.  The numerical values for the finite-size regime are the security parameter $\epsilon=10^{-9}$, and the discretization parameter $d=5$. The other parameters are Bob's detector efficiency $\eta_B = 0.61$, electronic noise $\nu_{B} = 0.12$, reconciliation efficiency $\beta=0.98$, and the excess noise $\xi=0.02$ (note that in a passive attack this noise is assumed to be a trusted preparation noise). The block size is chosen to be $N=10^{10}$, half of which is used in total for the parameter estimation. The modulation variance in the coherent-state protocol is optimized to maximise the key rate. We consider a probability  distribution for the free-space channel given by the elliptic-beam model \cite{Vasylyev.et.al.PRL.16}, where the model parameters have been chosen according to \cite{Neda-Nathan-Tim-freespace}.  }\label{keyrate}
\end{center}
\end{figure}

\begin{figure}[h]
    \begin{center}
      {\includegraphics[width=3.4 in, height=2.7 in] {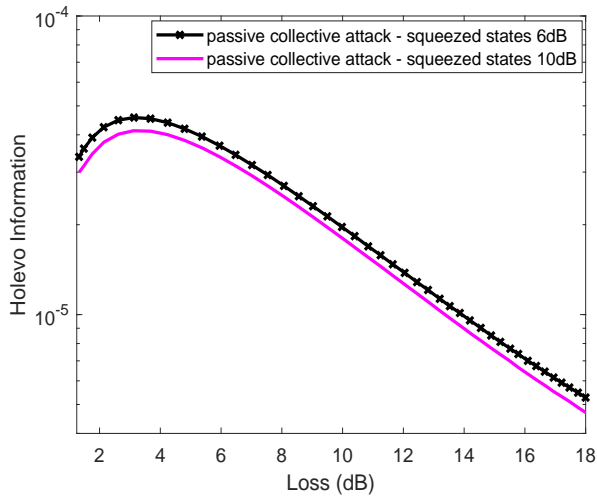}}
   \caption{Holevo information $\chi^{\epsilon_{\rm PE}}(b{:}E)$ resulting from the passive collective attack  as a function of channel loss (dB) for the non-ideal squeezed-state protocol with preparation noise $\xi$ (on top of the shot-noise variance) for 6 dB (black line) and 10 dB (magenta line) squeezing. The other parameters are the same as Fig.~\ref{keyrate}. 
     }\label{Holevo}
\end{center}
\end{figure}

\section{Conclusions}

We performed composable finite-size security analysis for a CV QKD protocol using an amplitude squeezed laser for free-space channels. Amplitude squeezing can be produced by compact semi-conductor lasers \cite{sqzlaser}. The information is encoded into the amplitude quadrature such that the Gaussian-modulated beam has the shot-noise variance, and detection is performed by the self-homodyne detection of the amplitude quadrature. Under the realistic assumption of classical monitoring of the line of sight, we limited the eavesdropper (Eve) to only passive attacks, where she can only collect the light lost in the communication. Under such an assumption, the shot-noise modulation  eliminates information leakage to Eve (and also eliminates the channel-fluctuation noise in the amplitude quadrature). As a result, the parameter estimation of Eve's information is no longer required. We investigated non-ideal modulation with some extra noise on top of the shot noise, which results in sufficiently small information leakage having negligible effect on the finite key rate. 
The protocol is highly robust to modulation noise, and can tolerate high values of channel loss. While our analysis shows the effectiveness of the protocol for losses up to 18 dB (given practical squeezing) for the block size of $10^{10}$, the performance can be improved by increasing the block size. For instance, for the block size of $10^{11}$, the protocol can tolerate losses up to 23 dB, expected in downlink channels from low-earth-orbit satellites. While our analysis focuses on Gaussian modulation, a remaining question would be how the performance is affected by (non-Gaussian) discrete modulation of squeezed states to the shot-noise level.

\section{Acknowledgments}
The authors gratefully acknowledge valuable discussions with Nathan Walk, Andrew Lance, and Thomas Symul.  
This research was supported by funding from the Australian Department of Defence. This research is also supported by the Australian Research Council (ARC) under the Centre of Excellence for Quantum Computation and Communication Technology (Project No. CE170100012).

\appendix

\section{Calculate Eve's Holevo information from the purification assumption}\label{AppendixA}

Let us consider a case where there is some preparation noise $\xi$ on top of the shot noise in Alice's side. In this case, the noise $\xi$ should be considered trusted in a passive attack scenario. This trusted noise is not attributed to  Eve, so it can be modeled by placing a beam splitter of transmissivity $\eta_p \to 1$ in Alice's side before the channel. The preparation noise can be modeled by a two-mode squeezed vacuum state, $\rho_{{F'_0}G'}$, of quadrature variance $\upsilon'   =  {\xi}/(1 - \eta_p )$. One input port of the beam splitter is the initial entangled mode $B_0$ with the $\hat q$ ($\hat p$) quadrature variance $b_q$ ($b_p$), and the second input port is fed by one half of the entangled state $\rho_{{F'_0}G'}$, mode~$F'_0$, while the output ports are mode~$B'_0$ (which is sent to Bob through the channel) and mode $F'$.

At the output of the channel Bob applies homodyne detection to the received mode~$B$. Bob's homodyne detector with efficiency $\eta_B$ and electronic noise variance of $\nu_B$ can be modeled by placing a beam splitter of transmissivity $\eta_B$ before an ideal homodyne detector. The homodyne detector's electronic noise can be modelled by a two-mode squeezed vacuum state, $\rho_{{F_0}G}$, of quadrature variance $\upsilon $, where $\upsilon  = 1 + {\nu_B}/(1 - \eta_B )$. One input port of the beam splitter is the received mode $B$, and the second input port is fed by one half of the entangled state $\rho_{{F_0}G}$, mode~$F_0$, while the output ports are mode~$B'$ (which is measured by the ideal homodyne detector) and mode $F$.

In a collective attack, Eve's information, $\chi(b{:}E)$, is given by $\chi(b{:}E)=\mathcal{S}(\rho_E)-\mathcal{S}(\rho_{E|B'})$. Here we assume Alice's preparation noise and Bob's detection noise are not accessible to Eve. In this case, the assumption that Alice and Bob's quantum state is purified by Eve's quantum state results in $\mathcal{S}(\rho_E) = \mathcal{S}(\rho_{AF'G'B})$, where the entropy $\mathcal{S}(\rho_{AF'G'B})$ can be calculated through the symplectic eigenvalues of covariance matrix ${\bf{M}}_{AF'G'B}$. The second entropy we require in order to determine $\chi(b{:}E)$ can be written as $\mathcal{S}(\rho_{E|B'})  = \mathcal{S}(\rho_{AF'G'FG|B'})$. The covariance matrix of the conditional state $\rho_{AF'G'FG|B'}$ is given by ${\bf{M}}_{AF'G'FG|B'}={\bf{M}}_{AF'G'FG}-{\bf{\sigma}}_{AF'G'FG,{B'}} \,\, ({\bf{X}} {\bf{M}}_{B'} {\bf{X}})^{\rm{ MP}} \,\, {\boldsymbol{\sigma}}^T_{AF'G'FG,{B'}}$.
Note that the matrices ${\bf{M}}_{AF'G'FG}, \boldsymbol{\sigma}_{AF'G'FG,{B'}}$, and ${\bf{M}}_{B'}$ can be derived from the decomposition of the covariance matrix
\begin{equation}\label{big-CM1}
{{\bf{M}}_{AF'G'FG{B'}}} = \left[ {\begin{array}{*{20}{c}}
   {{{\bf{M}}_{AF'G'FG}}} & {\boldsymbol{\sigma} _{AF'G'FG,{B'}}}  \\
   {{\boldsymbol{\sigma}^T_{AF'G'FG,{B'}}}} & {{{\bf{M}}_{{B'}}}}  \\
\end{array}} \right].
\end{equation}
Note that the covariance matrix ${\bf{M}}_{AF'G'FG{B'}}$ is given by the rearrangement of the following matrix  
\begin{equation}\label{big-CM2}
\begin{array}{l}
{\bf{M}}_{AF'G'{B'}FG} = 
({{\bf{I}}_A {\oplus} {\bf{I}}_{F'} {\oplus} {\bf{I}}_{G'} {\oplus} {\bf{S}}_{\rm bs} {\oplus} {\bf{I}}_G})^T \times
\\
\\
({\bf{M}}_{AF'G'{B}}  {\oplus} {\bf{M}}_{{F_0}G}) 
({{\bf{I}}_A  {\oplus} {\bf{I}}_{F'} {\oplus} {\bf{I}}_{G'} {\oplus} {\bf{S}}_{\rm bs} {\oplus} {\bf{I}}_G}),
\end{array}
\end{equation}
where ${\bf{S}}_{\rm bs}$ is the matrix for the beam splitter transformation (applied on modes $B$ and $F_0$), given by
\begin{equation}\label{BS-CMd}
{\bf{S}}_{\rm bs}= \left[ {\begin{array}{*{20}{c}}
{\sqrt{\eta_B}\,\bf{I}}&{\sqrt {1 - \eta_B} \,\bf{I}}\\
{-\sqrt {1 - \eta_B} \,\bf{I}}&{\sqrt{\eta_B}\,\bf{I}}
\end{array}} \right] ,
\end{equation}
and the covariance matrix of the entangled state $\rho_{{F_0}G}$ is given by
\begin{equation}\label{F0G-CM}
{\bf{M}}_{{F_0}G}= \left[ {\begin{array}{*{20}{c}}
{\upsilon\,\bf{I}}&{\sqrt {{\upsilon^2} - 1} \,\bf{Z}}\\
{\sqrt {{\upsilon^2} - 1} \,\bf{Z}}&{\upsilon\,\bf{I}}
\end{array}} \right] .
\end{equation}
Note that the covariance matrix ${\bf{M}}_{AF'G'{B}}$ is obtained by tracing out Eve's mode $E$ from the covariance matrix ${\bf{M}}_{AF'G'{B}E}$, given by  
\begin{equation}\label{big-CM3}
\begin{array}{l}
{\bf{M}}_{AF'G'{B}E} {=} ({{\bf{I}}_A {\oplus} {\bf{I}}_{F'} {\oplus} {\bf{I}}_{G'} {\oplus} {\bf{S}}^c_{\rm bs} })^T \times
\\
\\
({\bf{M}}_{AF'G'{B'_0}}  {\oplus} {\bf{M}}_{{E_0}}) 
({{\bf{I}}_A  {\oplus} {\bf{I}}_{F'} {\oplus} {\bf{I}}_{G'} {\oplus} {\bf{S}}^c_{\rm bs} }),
\end{array}
\end{equation}
where ${\bf{S}}^c_{\rm bs}$ is the matrix for the beam splitter (i.e. channel) transformation (applied on modes $B'_0$ and $E_0$), given by
\begin{equation}\label{BS-CMp}
{\bf{S}}^c_{\rm bs}= \left[ {\begin{array}{*{20}{c}}
{\sqrt{\eta_f}\,\bf{I}}&{\sqrt {1 - \eta_f} \,\bf{I}}\\
{-\sqrt {1 - \eta_f} \,\bf{I}}&{\sqrt{\eta_f}\,\bf{I}}
\end{array}} \right] ,
\end{equation}
where $\eta_f$ is the effective transmissivity of the free-space channel, and ${\bf{M}}_{{E_0}}$ is the covariance matrix of the vacuum state.  
Note that the covariance matrix ${\bf{M}}_{AF'G'{B'_0}}$ is given by the rearrangement of the following covariance matrix 
\begin{equation}\label{big-CM3}
\begin{array}{l}
{\bf{M}}_{A{B'_0}F'G'} {=} ({{\bf{I}}_A {\oplus} {\bf{S}'}_{\rm bs} {\oplus} {\bf{I}}_{G'}})^T \times \\
\\
({\bf{M}}_{A{B_0}}  {\oplus} {\bf{M}}_{{F'_0}G'} )({{\bf{I}}_A {\oplus} {\bf{S}'}_{\rm bs} {\oplus} {\bf{I}}_{G'}}),
\end{array}
\end{equation}
where ${\bf{S}'}_{\rm bs}$ is the matrix for the beam splitter transformation (applied on modes $B_0$ and $F'_0$), which is given by
\begin{equation}\label{BS-CMp}
{\bf{S}'}_{\rm bs}= \left[ {\begin{array}{*{20}{c}}
{\sqrt{\eta_p}\,\bf{I}}&{\sqrt {1 - \eta_p} \,\bf{I}}\\
{-\sqrt {1 - \eta_p} \,\bf{I}}&{\sqrt{\eta_p}\,\bf{I}}
\end{array}} \right] .
\end{equation}
Note that the covariance matrix of the entangled state $\rho_{{F'_0}G'}$ is given by 
\begin{equation}\label{F'0G'-CM}
{\bf{M}}_{{F'_0}G'}= \left[ {\begin{array}{*{20}{c}}
{\upsilon'\,\bf{I}}&{\sqrt {{\upsilon'^2} - 1} \,\bf{Z}}\\
{\sqrt {{\upsilon'^2} - 1} \,\bf{Z}}&{\upsilon'\,\bf{I}}
\end{array}} \right] ,
\end{equation}
and the covariance matrix ${\bf{M}}_{A{B_0}}$ is given by Eq.~(\ref{sqz-initial-state-Gaussian}). Note that with the assumption of zero (trusted) fluctuation noise, the purification of Alice and Bob's state by Eve's state results in ${\bf{M}}_E = {\bf{M}}_{AF'G'{B}}$, and ${\bf{M}}_{E|B'} = {\bf{M}}_{AF'G'FG|B'}$. Therefore, Eve's information, $\chi(b{:}E)$,  calculated from Eve and Bob's covariance matrix (discussed in the main text) is the same as that calculated based on the purification assumption (discussed above).

\section{Parameter estimation for squeezed-state protocol}\label{AppendixB}

In the prepare-and-measure scheme, for a sub-channel with transmissivity $\eta$, we can consider a normal linear model for Alice and Bob's correlated $q$ quadrature variables, $q_{A}$ and $q_{B}$, respectively,
\begin{equation}\label{eqn-subchannel}
q_{B} = t_s  q_{A} + q_{n,s},
\end{equation}
where $t_s = \sqrt{{\eta_B \eta}} $, and $q_{n,s}$ follows a centred normal distribution whose variance is determined from the observed data as follows, $\sigma_s^2 = 1 + \nu_B +  {\eta_B} \eta (V_{\rm sqz} + {\xi}) - {\eta_B} \eta $ (note that Alice's variable $q_{A}$ has the variance $V_{\rm sig}$). Using the revealed data of size $k_s$ for the sub-channel (note in our numerical simulation we assumed $10^5$ sub-channels), the maximum-likelihood estimators for the sub-channel parameters, $t_s$ and $\sigma^2_s$, are given by \cite{MLE-estimator2010,MLE-estimator2012}
\begin{equation}\label{t-subchannel}
\begin{array}{l}
 \hat t_s = \frac{{\sum\nolimits_{i = 1}^{k_s} {{{A_{i}}}{{B_i}}} }}{{\sum\nolimits_{i = 1}^{k_s} {{{{{A_i^2}}}}} }}, \\
 \\
 {\hat \sigma^2_s} = \frac{1}{{k_s}}\sum\nolimits_{i = 1}^{k_s} {{{({{B_{i}}} - \hat t_s {{A_{i}}})}^2}},
 \end{array}
\end{equation}
where $A_{i}$ and $B_{i}$ are the realizations of  $q_{A}$ and $q_{B}$ for the sub-channel, respectively. The confidence interval for $t_s$ is given by $t_s \in [\hat t_s-\Delta (t_s),\hat t_s+\Delta (t_s) ]$, where
\begin{equation}\label{t-ci-subchannel}
\begin{array}{l}
\Delta (t_s) = {z_{\epsilon_{\rm PE} /2}}\sqrt {\frac{{{\hat \sigma ^2_s}}}{k_s V_{\rm sig}}}.
 \end{array}
\end{equation}
The estimator of the square root of sub-channel transmissivity, and its error bar is given by
\begin{equation}\label{eta-ci-subchannel}
\begin{array}{l}
\widehat{{ \sqrt{\eta}}} = \frac{   \hat t_s  }{{{\sqrt{\hat \eta_B}}}  },\\
\\
\Delta ({\sqrt{\eta}}) = \widehat{{ \sqrt{\eta}}} \sqrt{ { \left| {\frac{{ {\Delta(t_s)}}}{{{\hat t_s}}}} \right|^2 + \left| {\frac{{\Delta (\eta_B )}}{{ 2\hat \eta_B }}} \right|^2 } },
 \end{array}
\end{equation}
Here, we generalize the above discussed parameter estimation method to the data of size $k$ revealed over all sub-channels to estimate $\xi$. Considering Eq.~(\ref{eqn-subchannel}) over all sub-channels we can still have a normal linear model for Alice and Bob's correlated $q$ quadrature variables as the following
\begin{equation}\label{eqn}
q_{B} = t  q_{A} + q_{n},
\end{equation}
where $t =  \sqrt{\eta_B \langle  \eta \rangle} $, and $q_{n}$ follows a centred normal distribution whose variance is determined from the observed data as follows, $\sigma^2 = 1 + \nu_B +  \eta_B \langle { \eta } \rangle (V_{\rm sqz} + {\xi}) - {\eta_B} \langle \eta \rangle $. Using the total data revealed  over all sub-channels of size $k$, we can calculate the maximum-likelihood estimators for $  t  $ and $  \sigma^2 $, which are given by
\begin{equation}\label{t}
\begin{array}{l}
\hat t = \frac{{\sum\nolimits_{i = 1}^{k} {{{A_{i}}}{{B_i}}} }}{{\sum\nolimits_{i = 1}^{k} {{{{{A_i^2}}}}} }}, \\
 \\
\hat  \sigma^2  = \frac{1}{{k}}\sum\nolimits_{i = 1}^{k} {{{({{B_{i}}} - \hat t {{A_{i}}})}^2}}.
 \end{array}
\end{equation}
The confidence intervals for these parameters are given by $t \in [\hat t-\Delta (t),\hat t+\Delta (t) ]$, and $ \sigma^2  \in [  \hat \sigma^2 -\Delta ( \sigma^2 ),\hat  \sigma^2 +\Delta ( \sigma^2 ) ]$ where
\begin{equation}\label{t-ci}
\begin{array}{l}
\Delta (t) = {z_{\epsilon_{\rm PE} /2}}\sqrt {\frac{{{  \hat \sigma ^2 }}}{k V_{\rm sig}}},  \\
\\
\Delta ({{\sigma }^2}) = {z_{\epsilon_{\rm PE} /2}}\frac{{{  \hat \sigma  ^2}\sqrt 2 }}{{\sqrt {k} }}.
 \end{array}
\end{equation}
Note that when no signal is exchanged, Bob's variable with realization $B_{0i}$ follows a centred normal distribution whose variance is determined from the observed data as follows, $\sigma_0^2 = 1 + \nu_B $, which is Bob's shot noise variance. The maximum-likelihood estimator for $\sigma_0^2$ is given by $\hat \sigma_0^2 = \frac{1}{{N}}\sum\nolimits_{i = 1}^{N} {{{{{B^2_{0i}}} }}}$. The confidence interval for this parameter is given by $\sigma_0^2 \in [\hat \sigma_0^2-\Delta (\sigma_0^2),\hat \sigma_0^2+\Delta (\sigma_0^2) ]$, where $\Delta (\sigma_0^2)={z_{\epsilon_{\rm PE} /2}}\frac{{{\hat \sigma_0 ^2}\sqrt 2 }}{{\sqrt {N} }} $ \footnote{Note that ${z_{\epsilon_{\rm PE} /2}}$ is such that $1-{\rm erf}(\frac{{z_{\epsilon_{\rm PE} /2}}}{\sqrt{2}})/2 = \epsilon_{\rm PE} /2$, where erf
is the error function.}.
Now we can estimate $\langle \eta \rangle$ and $\xi$, which are given by
\begin{equation}\label{eta-average-ci}
\begin{array}{l}
\widehat{\langle  \eta \rangle}  = \frac{  \hat t ^2 }{{{\hat \eta_B}}  },\\
\\
\Delta (\langle \eta \rangle) =  \widehat{\langle  \eta \rangle} \sqrt{ { \left| {\frac{{ 2{\Delta( t )}}}{{{\hat  t }}}} \right|^2 + \left| {\frac{{\Delta (\eta_B )}}{{ \hat \eta_B }}} \right|^2 } },
\\
\\
\widehat  \xi  =  \frac{\hat  \sigma^2  -\hat \sigma_0^2}{  \hat \eta_B \widehat{\langle  \eta \rangle}} - \hat V_{\rm sqz} + 1 ,\\
\\
\Delta (\xi) = 
\\
\widehat{\xi}  \sqrt{ {\left| {\frac{{\Delta ({ \sigma ^2 })}}{{{\hat \sigma^2 } {-} \hat \sigma _0^2}}} \right|^2 {+} \left| {\frac{{\Delta (\sigma _0^2)}}{{{  \hat \sigma^2 } {-} \hat \sigma _0^2}}} \right|^2 {+} \left| {\frac{{\Delta (\eta_B )}}{{\hat \eta_B }}} \right|^2 {+} \left| {\frac{{\Delta (\langle \eta \rangle )}}{{\widehat{\langle  \eta \rangle} }}} \right|^2} } {+} \Delta(V_{\rm sqz}).
\end{array}
\end{equation}
Note that in order to maximise Eve's information, $\chi^{\epsilon_{\rm PE}}(b{:}E)$, from passive collective attacks, the worst-case estimators of $\eta$  and $\xi $ should be used to evaluate Eve's information. Now, having Eqs.~(\ref{eta-ci-subchannel}) and (\ref{eta-average-ci}), the worst-case estimators of  parameters $\sqrt{\eta}$ (for channel losses above 3 dB) to calculate $\eta_f$ and $\xi $ are given by $\widehat{{ \sqrt{\eta}}}+\Delta(\sqrt{\eta})$  and $\hat \xi+\Delta(\xi)$. 
Note also that for the parameter estimation of the coherent-state protocol, we can still use the equations provided in this section, however we should set $V_{\rm sqz}=1$ due to the use of coherent states.

Note that here we estimated the noise $\xi$ by first estimating the channel transmissivity and revealing a subset of data. However, since the preparation noise $\xi$ is trusted and not under Eve's control, it can be estimated using the whole data block with a very good precision. This estimation can be performed in only Alice's lab without revealing any data. 

\end{document}